\documentclass[english,twoside,a4paper,10pt]{article}
\usepackage[utf8]{inputenc}
\usepackage[T1]{fontenc}
\usepackage{amsmath}
\usepackage{amsfonts}
\usepackage{graphicx}
\usepackage{subfigure}
\usepackage{a4wide}
\usepackage{amssymb}
\usepackage{fancyhdr}
\usepackage{mathrsfs}
\usepackage{amssymb}
\usepackage[toc,page]{appendix}
\usepackage{xcolor}
\usepackage{float}
\usepackage{cite}
\usepackage{hyperref}
\usepackage{fontawesome}
\usepackage[dvipsnames]{xcolor}
\usepackage{authblk}

\def\ba{\begin{eqnarray}}
\def\ea{\end{eqnarray}}

\def\be{\begin{equation}}
\def\ee{\end{equation}}

\begin{document}

\title{\centering{$\tt GrayHawk$ $\tt v2$: wormholes and numeric extension}}

\author[1,2]{Marco Calz\'a}
\affil[1]{Dipartimento di Fisica, Universit\`a di Trento,
Via Sommarive 14, 38123 Povo (TN), Italy}
\affil[2]{Univ Coimbra, Faculdade de Ci\^encias e Tecnologia,
da Universidade de Coimbra and CFisUC,
Rua Larga, 3004-516 Coimbra, Portugal}

\date{}

\maketitle
\begin{abstract}
    We enlarged the capabilities of the publicly available Mathematica-based tool {\tt GrayHawk}. This second version expands the spectrum of metrics that can be considered in two distinct and disjoined directions. First, it enables a fully numerical computation of the tortoise coordinates integral, allowing the user to account for many metrics for which an analytic computation was impractical. Second, it extends the scattering problem to wormhole solutions. The pool of pre-loaded metrics is enriched, enabling immediate testing of the new features, and the code's modular structure is maintained to facilitate easy modification. This implementation proves {\tt GrayHawk} adaptability and makes it an even more powerful tool for studying black holes, wormholes, Hawking radiation, and other features involving field propagation on curved manifolds. The codes described in this work are publicly available at \href{https://github.com/marcocalza89/GrayHawk-v2}{\faGithub}.
\end{abstract}

\section{Introduction}

Black holes remain among the most compelling predictions of general relativity, representing the strongest known manifestation of gravitation and providing a natural arena in which classical and quantum effects intersect. Their relevance spans astrophysics, cosmology, and fundamental theory, from the dynamics of compact-object mergers detected through gravitational waves to the long-standing problem of information loss associated with Hawking evaporation \cite{Hawking:1975vcx,Page:1977um,Page:1976ki,Page:1976df,Unruh:1977ga}. In recent years, precision observations by the LIGO-Virgo-KAGRA collaboration and by the Event Horizon Telescope have transformed black holes from purely theoretical constructs into directly testable laboratories of strong-field gravity \cite{LIGOScientific:2016lio,LIGOScientific:2018dkp,LIGOScientific:2021sio, EventHorizonTelescope:2019dse,EventHorizonTelescope:2021dqv,EventHorizonTelescope:2022xqj,EventHorizonTelescope:2022wkp}.  

\noindent A key quantity governing several observable and semiclassical phenomena is the set of \emph{gray-body factors}, namely the frequency-dependent transmission coefficients associated with wave propagation on curved backgrounds. These coefficients encode the probability that a field mode incoming from infinity penetrates the effective potential barrier surrounding a compact object. They therefore determine the departure of Hawking radiation from an ideal blackbody spectrum and play a central role in absorption cross sections, scattering amplitudes, and energy emission rates \cite{Teukolsky:1972my,Teukolsky:1973ddt,Teukolsky:1973ha,Teukolsky:1974yv,Page:1977um,Page:1976ki,Page:1976df}. In static and spherically symmetric geometries, the perturbation equations for fields of spin $s=0,\frac12,1,\frac32,2$ reduce to Schr\"odinger-like radial equations, revealing the associated transmission and reflection coefficients. Reliable numerical determination of these quantities is therefore essential for a broad range of applications.

\noindent Motivated by these considerations, Ref.~\cite{Calza:2025whq} introduced the publicly available Mathematica-based package \texttt{GrayHawk v2}, designed to compute gray-body factors for massless fields propagating on spherically symmetric and asymptotically flat black-hole spacetimes. Its modular structure and pre-loaded set of metrics were conceived to provide a flexible tool for both phenomenological studies and exploratory model building. \texttt{GrayHawk} versatility was at the base of several studies\cite{Han:2026fpn,Calza:2025yfm,Han:2025cal,Calza:2025mwn,Yuan:2025eyi,Pedrotti:2025idg,Arbey:2025dnc} and this it has been used to implement \texttt{BlackHawk v3.0} \cite{Arbey:2026koc} a publicaly available code released together with \texttt{GrayHawk}. \texttt{BlackHawk v3.0} is the latest release of the \texttt{BlackHawk} code \cite{Arbey:2019mbc,Auffinger:2020ztk,Arbey:2021mbl,Auffinger:2022sqj} which expands the pool of available metrics, improves the computation in the case of Kerr metric, and has more performing extrapolation routines. However, ongoing interest in non-standard compact objects and in metrics lacking closed-form tortoise coordinates naturally calls for a broader computational framework.

\noindent Among the most studied alternatives to classical black holes are traversable wormholes. Originally formalized in the modern context by Morris and Thorne \cite{Morris:1988tu}. In the context of general relativity, wormholes connect two asymptotically flat regions through a throat sustained by matter violating the standard energy conditions. Although no observational evidence currently supports their existence, wormholes continue to play an important role in gravitational theory, semiclassical gravity, and quantum-information-inspired approaches to spacetime structure \cite{Visser:1995cc,Cardoso:2019rvt,Mondal:2025tht}. Their phenomenology is especially intriguing because, unlike black holes, they possess no event horizon. As a consequence, incoming radiation is not irreversibly absorbed but can traverse the throat and partially re-emerge, generating distinctive scattering signatures.

\noindent One of the most widely discussed manifestations of this horizon-less structure is the appearance of late-time echoes in ring-down signals. In black-hole spacetimes, perturbations generated after a merger are largely absorbed at the horizon after exciting the photon-sphere barrier. By contrast, in wormhole geometries, the existence of two potential barriers or of an internal reflective cavity allows part of the wave-packet to bounce repeatedly, producing a train of delayed pulses \cite{Cardoso:2016oxy,Bueno:2017hyj,Yang:2021cvh}. If the effective cavity extends between two turning points, the characteristic time delay between successive echoes is approximately dependent on the distance between the points.
Hence, the echo spacing directly probes the geometry through the tortoise coordinate.

\noindent The scattering origin of echoes is extensively studied, and it is demonstrated that wave-packets evolving on a wormhole background produce echoes. It was shown that significant echo amplitudes occur only within specific frequency bands and that the cavity structure can also affect polarization properties of gravitational perturbations. More recently, renewed attention has been devoted to wormhole ringdown in increasingly realistic settings, including black-bounce geometries \cite{Ou:2021efv,Yang:2021cvh}, asymmetric wormholes \cite{Magalhaes:2023har,Guendelman:2008zp,Guendelman:2009pf,Guendelman:2009zd,Guendelman:2009zz}, thin-shell constructions \cite{Poisson:1995sv,Visser:1989kg,Liempi:2024yjd,Sarkar:2026pjg,Eid:2024iza,Guo:2022iiy,Kokubu:2020lxs}, and traversable models supported by exotic quantum matter. These studies reinforce the view that transmission and reflection coefficients are not merely auxiliary quantities, but fundamental observables linking the internal geometry to potentially measurable signals.

\noindent From a computational perspective, both black holes and wormholes pose analogous wave-scattering problems. In each case, one must determine the effective potential, construct the tortoise coordinate, impose appropriate boundary conditions, and extract transmission amplitudes over wide frequency ranges. However, often the tortoise integral cannot be expressed analytically, and a robust numerical treatment of the tortoise coordinate is therefore a necessary step toward a unified platform for compact-object scattering.

\noindent In this work, we present \texttt{GrayHawk} $\tt v2$, which substantially enlarges the scope of the original package in two complementary directions. First, we implement a fully numerical evaluation of the tortoise-coordinate integral, thereby allowing the treatment of metrics for which analytic expressions are unavailable or impractical. Second, we extend the scattering module to traversable wormholes, enabling the computation of transmission coefficients and related observables for horizonless compact objects. The pool of pre-loaded metrics is correspondingly expanded, while preserving the modular philosophy of the original code so that users may easily implement additional geometries.

\noindent This new release demonstrates the adaptability of \texttt{GrayHawk} and provides a versatile tool for the study of wave propagation, gray-body factors, Hawking emission, and echo phenomenology in both black-hole and wormhole spacetimes.

\noindent This paper is organized as follows: in Section 2, we outline the theoretical formalism, introducing and describing the scattering problem in the case of black holes and wormholes. In Section 3, we report and quickly review the newly introduced metrics. Section 4 is devoted to the description of the structure of the tool, its numerical methods and algorithms implemented, accuracy, efficiency, and weaknesses. In Section 6, we validate the newly introduced fully numeric tortoise computation by applying it to wormhole metrics and reproducing results present in the literature. Finally, conclusions are drawn in Section 6. In this work, we set $c=\hbar=G=1$.

\section{Theoretical formalism} \label{NGH}
As in the case of \cite{Calza:2025whq}, we focus on the subset of Petrov type D \cite{Petrov:2000bs} metrics described by spherically-symmetric static metrics whose general form in four-dimensional Boyer-Lindquist coordinates reads
\begin{equation}\label{eq:metric}
	 ds^2=-G(r) dt^2+\frac{dr^2}{F(r)} +H(r) d\Omega^2\;,
\end{equation}
with $d\Omega^2=d\theta^2+\sin^2(\theta)d\varphi^2$ the solid angle infinitesimal element in spherical coordinates. As for \cite{Calza:2025whq}, we additionally require asymptotically flatness
\begin{equation}\label{falloffs}
	F(r)\underset{r\rightarrow+\infty}{\longrightarrow}1\;,\;\;\;\; G(r)\underset{r\rightarrow+\infty}{\longrightarrow}1\;,\;\;\;\; H(r)\underset{r\rightarrow+\infty}{\sim}r^2\,.
\end{equation}
Many Black Holes (BHs) and WormHoles (WHs) solutions fit this description, from the most classical GR vacuum solutions to the ones emerging from non-linear quantum electrodynamics, Loop Quantum Gravity (LQG), extra dimensions, and so on, to more exotic phenomenologically inspired metrics.

\noindent As better described in \cite{Calza:2025whq}, with the use of the Newman-Penrose (NP) formalism it is possible to condense all the spin $s=0,1/2, 1,3/2, 2 $ massless equations into a single master equation, which under the Ansatz of separability of variables can be decomposed into the spin weighted spherical harmonics equation satisfying \cite{Chandrasekhar1990,Kalnins:1992,Fackerell&Crossman1977,Suffern1983,Seidel:1988ue,Berti:2005gp}
\begin{equation}
	\left(\frac{1}{\sin\theta}\partial_\theta(\sin\theta\,\partial_\theta)+\csc^2\theta\,\partial_\varphi^2+\frac{2is\cot\theta}{\sin\theta}\partial_\varphi+s-s^2\cot^2\theta+\lambda_l^s\right)S_{l,m}^s=0\,,
\end{equation}
with $\lambda_l^s= l(l+1)-s(s+1)$ separation constant, and a radial part satisfying
\begin{equation}\label{eq:teukolsky_general}
	A_s\big(B_s\Phi'_s\big)'+\left(\frac{H}{G}\omega^2+i\omega s\sqrt{\frac{F}{G}}\left(H'-H \frac{G'}{G}\right)+C_s\right)\Phi_s=0\,,
\end{equation}
where
\begin{eqnarray}
A_s= \sqrt{\frac{F}{G}} \frac{1}{(G H)^s}\,,
\label{eq:as}
\end{eqnarray}
\begin{eqnarray}
B_s=\sqrt{F G } (G H)^s H\,,
\end{eqnarray}
\begin{align}
C_s &= s \frac{F H G''}{G} + \frac{s}{2} \left ( \frac{H F' G' }{G} - \frac{F H G'^2}{G^2} \right ) \nonumber \\ 
&+ \frac{s(3-2s)}{4} \left( 2 F H'' + F' H'  \right) +\frac{s(2s-1)}{4} \frac{F H'^2}{H}\nonumber \\ 
&+\frac{s(2s+5)}{4}\frac{F G' H'}{G}-\lambda^s_l-2s\,.
\end{align}
The radial equation can be rewritten in the Schr\"odinger-like form
\begin{equation} \label{SchEq}
    \partial^2_{r^*}Z_s + \Bigl(\omega^2-V_s(r(r^*))\Bigr)Z_s =0
\end{equation}
where $r^*$ are the tortoise coordinates defined by
\begin{equation}\label{tort}
    \frac{dr^*}{dr} = \frac{1}{\sqrt{F G}}
\end{equation}
and
\begin{subequations}\label{eq:potentials}
\begin{align}
	&V_0=\nu^0_l\frac{G}{H}+\frac{\partial_{r^*}^2\sqrt{H}}{\sqrt{H}}\,,\\
	&V_1=\nu^1_l\frac{G}{H}\,,\\
	&V_2=\nu^2_l\frac{G}{H}+\frac{(\partial_{r^*}H)^2}{2H^2}-\frac{\partial_{r^*}^2\sqrt{H}}{\sqrt{H}}\,,\\
	&V_{1/2}=\nu^{1/2}_l\frac{G}{H} \pm \sqrt{\nu^{1/2}_l}\,\partial_{r^*}\left( \sqrt{\frac{G}{H}} \right)\,,
\end{align}
\end{subequations}
and $\nu^s_l=l(l+1)-s(s-1)$.\\
\subsection{Black Holes}
\noindent In the case of BHs, the metric has an event horizon at $r=r_H$, namely a simultaneous root of $F$ and $G$, and is asymptotically flat for $r \rightarrow + \infty$. 
Therefore, $V_s$ vanishes at the horizon ($r^* \rightarrow -\infty$) and at infinity ($r^* \rightarrow +\infty$). This way, the asymptotic solutions read
\begin{subequations}
\begin{align}
	&Z_s(r^*\rightarrow -\infty) = \mathfrak{a} \;e^{i \omega r^*}+\mathfrak{b} \;e^{-i \omega r^*} \label{asympt-} \,,\\
    &Z_s(r^*\rightarrow +\infty) = a \;e^{i \omega r^*}+b \;e^{-i \omega r^*} \label{asympt+}\,.
\end{align}
\end{subequations}
On the event horizon, we invoke purely in-going boundary conditions and normalize the wave function to unity. Namely, $\mathfrak{a}=0$ and $\mathfrak{b}=1$.

\noindent Under such choices, we can integrate out the solution to large radial distances using Eq.~(\ref{SchEq}) and solve the scattering problem, obtaining the transmission and reflection coefficients
\begin{equation}\label{RT}
    R=\frac{1}{|a|^2}\;\;\;T=\frac{1}{|b|^2}\;.
\end{equation}
Those coefficients are functions of the energy and depend on the field spin and mode.\\

\noindent The careful reader may notice that integrating Eq.~(\ref{tort}) explicitly to have the functional form of $r^*(r)$ leaves an arbitrary integration constant to fix. Namely,
\begin{equation}\label{expltort}
  r^*(r)= I(r) + C_1
\end{equation}
with $I(r)$ continuous function such that $ \lim\limits_{r \rightarrow r_H} =-\infty $ and $ \lim\limits_{r \rightarrow +\infty} =r $. 

\noindent Therefore, whatever choice of the constant $C_1$ will provide the same asymptotic behaviors of $r^*(r)$, giving a well-behaved $V_s$ and the same plane wave approximation at the horizon and at infinity. For this reason, and for simplicity, we choose $C_1=0$.\\ 

\noindent The number of particles of a given species $i$ with spin $s$, emitted by Hawking radiation, per unit time and energy, would be thermally distributed in the absence of the scattering problem described by Eq.~(\ref{SchEq})~\cite{Hawking:1975vcx,Page:1976df,Page:1976ki,Page:1977um}. The transmission coefficient $T$, in this case, acts as a filter of an otherwise black-body radiation, and it is therefore historically referred to as the Gray-Body Factor (GBF) and denoted with $\Gamma_s^l(\omega)$. Therefore, GBFs play an important role in the study of BH particle-emission phenomenology, including beyond-the-standard-model, and Regular BHs (RBHs) \cite{Calza:2021czr,Pedrotti:2024znu,Calza:2022ljw,Calza:2023rjt,Calza:2023gws,Calza:2023iqa,Calza:2025mrt,Calza:2025yfm,Calza:2022ioe}.\\

\noindent The Hawking radiated primary spectrum therefore reads:~\footnote{This expression implicitly assumes that the particles emitted by the BH are not coupled to the regularizing parameter $\ell$, a reasonable assumption.}
\begin{eqnarray}\label{prim}
\frac{d^2N_i}{dtdE_i}=\frac{1}{2\pi}\sum_{l,m}\frac{n_i\Gamma^s_{l,m}(\omega)}{ e^{\omega/T}\pm 1}\,,
\label{eq:d2ndtdei}
\end{eqnarray}
with the plus (minus) sign in the denominator associated to fermions (bosons), and where $n_i$ is the number of degrees of freedom of the particle in question, $\omega=E_i$ is the mode frequency (in natural units). We implicitly set $k_B=1$.
Finally, the temperature is related to the surface gravity computed at the horizon and reads
\begin{eqnarray}
T=\sqrt{\frac{F(r)}{G(r)}}\frac{G'(r)}{4\pi}\vert_{r_H}\,,
\label{eq:temperature}
\end{eqnarray}

\subsection{Wormholes}
WHs metrics are characterized by the presence of a throat at $r=r_T$ connecting distinct parts of a manyfold. In an extrinsic description, such as the one predominantly used in this paper and in \cite{Calza:2025whq}, this means that there will be values of the coordinates set $(t,r,\theta,\varphi)$ corresponding to no point on the manifold, values corresponding to 2 points on the manifold, one on one side of the WH one on the other. Oftentimes, the two sides of the WH are labeled as up and down universes. It is possible to adopt an intrinsic description by the introduction of the proper distance defined by
\begin{equation}
    \mathfrak{l}(r)=\pm\int_{r_T}^r \frac{dr'}{F(r')}
\end{equation}
In this case, the sign of $\mathfrak{l}$ tells in which universe one is, on the throat $\mathfrak{l}=0$, and all values of the coordinate system $(t,\mathfrak{l},\theta,\varphi)$ correspond to a single point on the manifold.

\noindent It is possible to distinguish between two primary classes of wormholes based on their geometric continuity: thin-shell and smooth wormholes. A thin-shell wormhole is typically constructed using a "cut-and-paste" technique, where two regions of spacetime (such as two Schwarzschild geometries) are joined at a common boundary; this results in a mathematical discontinuity at the throat, implying that all supporting exotic matter is concentrated on a singular, infinitely thin surface. In contrast, a smooth wormhole, exemplified by the Ellis-Bronnikov metric, an example of WH of the Morris and Thorn type, features a curvature and stress-energy tensor that vary continuously throughout the transition. In these models, the exotic matter is not a surface layer but a distributed fluid or field, making the geometry twice-differentiable everywhere and avoiding the use of junction conditions.

\noindent For continuity with the previous choices of formalism, we will adopt the extrinsic description, having in mind the double (therefore non-functional) correspondence between coordinates and manifold points. We may notice that, in such a case, Eq.~(\ref{expltort}) describes, through a function, the behavior of the tortoise coordinate in one universe. We will consider WHs symmetric, in which the two universes are identical. Therefore, it is possible to achieve the full tortoise coordinate by adding to $I(r)$ its reflection with respect to the horizontal line $I(r_T)$. We decide to take the degree of freedom $C_1=-I(r_T)$ in such a way that $r^*(r_T)=0$. This way, the WH throat in tortoise coordinates is located in $r^*=0$ and $r^*>0$ corresponds to one universe, while $r^*<0$ to the other. We underline that, as in the case of BHs, this choice do not affect the well behavior of $V_s$ vanishing far away from the throat for both universes at $r^* \rightarrow \pm\infty$. Therefore, the asymptotic solutions to Eq.~(\ref{SchEq}) read as ~(\ref{asympt-})~(\ref{asympt+}), not affecting the scattering problem we will consider.

\noindent We aim to study the well-known problem of echoes, which characterize the scattering of a wave on a WH, and analyze the propagation of a wave from one universe through the other as it passes the throat of a WH. Therefore, we impose $\mathfrak{a}=0$ and $\mathfrak{b}=1$, integrate the solution to the other universe using Eq.~(\ref{SchEq}) having again (\ref{RT}).

\section{New metrics}

Here we report the metrics that the user may find pre-loaded in $\tt GrayHawk$ $\tt v2$. We note that in the previous version, the Reisner-Nordstr\"om metric was provided as an example of how to modify the code. The philosophy of code modifications expressed in that example remains valid, but now the Reisner-Nordstr\"om metric is among the preloaded metrics.
\subsection{Culetu-Ghosh-Simpson-Visser Black Hole}
The regular space-times analyzed in the previus version are characterized by de Sitter (dS) cores, a feature commonly encountered in several RBH metrics. An alternative and phenomenologically compelling scenario involves so-called “hollow” RBHs, in which the central singularity is replaced by an asymptotically Minkowski core. In this case, both the energy density and pressure vanish asymptotically toward the center. This behavior contrasts sharply with that of dS cores, where the energy density approaches a finite value associated with a positive cosmological constant, and the pressure tends to an equal magnitude with opposite sign.

\noindent From a theoretical and mathematical perspective, RBHs with Minkowski cores present several appealing features. In particular, the vanishing energy density considerably simplifies the physical description in the deep core region. Moreover, the typically cumbersome solutions of polynomial equations—often not expressible in closed form—can be replaced by more tractable special functions, thereby enhancing the analytical manageability of the space-time. From a physical standpoint, our motivation for investigating this class of BHs is to extend the range of physical properties and phenomenological implications of PRBHs beyond the dS-core configurations explored to date.

\noindent Motivated by these considerations, we examine an RBH solution with a Minkowski core independently investigated by Culetu~\cite{Culetu:2013fsa,Culetu:2014lca}, Ghosh~\cite{Ghosh:2014pba}, and Simpson and Visser~\cite{Simpson:2019mud}. Although no standard nomenclature exists in the literature, we shall refer to this solution as the CGSV BH, following the initials of the aforementioned authors. The corresponding space-time is defined by the metric function:
\begin{eqnarray}
F(r)=G(r) = 1-\frac{2M}{r}\exp \left ( -\frac{\ell}{r} \right )\;\;,\;\;\;H(r)=r^2.
\label{eq:frcgsv}
\end{eqnarray}
In contrast to the Bardeen and Hayward cases, the horizon radius $r_H$ admits a closed-form expression given by:
\begin{eqnarray}
r_H=-\frac{\ell}{W \left ( -\frac{\ell}{2 M} \right ) },
\label{eq:rhcgsv}
\end{eqnarray}
where $W$ denotes the Lambert function. Restricting to the principal branch $W_0$, a real and positive horizon radius exists provided that:
\begin{eqnarray}
W_0 \left ( -\frac{\ell}{2 M} \right ) \leq 0 \implies 0 \leq \ell < \frac{2M}{e},
\label{eq:lambhert}
\end{eqnarray}
or equivalently $0 \leq \ell < r_H$. Although the CGSV BH was originally introduced on phenomenological and mathematical grounds, it has been demonstrated in Refs.~\cite{Kumar:2020ltt,Singh:2022xgi} that this geometry can arise within General Relativity (GR) coupled to a suitable non-linear electrodynamics source. In such a framework, denoting by $g$ the non-linear electrodynamics coupling constant (or charge), the regularization parameter is given by $\ell = g^2/2M$, where $M$ is the BH mass. Nonetheless, in line with the treatment of other RBHs considered in this work, we regard the CGSV solution as a toy model describing a regular space-time endowed with a Minkowski core.

\noindent This metric was explicitly mentioned in the paper, complementing the previous version of this code \cite{Calza:2025whq}. In fact, this metric exemplifies a situation in which the tortoise's coordinates cannot be obtained in an analytical way using {\tt Integrate[]}. This remains true, but it is now possible to obtain the GBFs for this BH using the fully numeric computation of the tortoise coordinates described in the next section.

\subsection{Dymnikova Black Hole}

The Dymnikova spacetime is a static, spherically symmetric, asymptotically flat regular black-hole solution in which the central curvature singularity is replaced by a de~Sitter core. It can be written as
\begin{equation}
F(r)=G(r)=1-\frac{2M}{r} \left(1-e^{-\,r^3/\ell^3}\right).
\end{equation}

\noindent Here, $M$ is the total ADM mass measured at spatial infinity. At the same time, $\ell$ is a length scale controlling the size of the regular core and related to the effective vacuum energy density at the origin.

\noindent For large radial distances, the exponential term vanishes, and the metric approaches the Schwarzschild geometry, while for $r\to 0$
\begin{equation}
F(r)\simeq 1-\frac{2M}{\ell^3}r^2,
\end{equation}
which is the behavior of a de~Sitter core. Consequently, all curvature invariants remain finite at $r=0$, and the spacetime is free of the Schwarzschild singularity.

\noindent Depending on the ratio between $M$ and $\ell$, the solution may admit two horizons (Cauchy and event horizon), one degenerate extremal horizon, or no horizons. In the two-horizon case, the Dymnikova geometry represents a regular black hole interpolating smoothly between an exterior Schwarzschild region and an interior de~Sitter vacuum core.

\noindent Because of its regularity and analytic simplicity, this metric is frequently employed in studies of black-hole thermodynamics, quasi-normal modes, Hawking radiation, and wave propagation on non-singular compact-object backgrounds.

\subsection{Zhang-Lewandowski-Ma-Yang Black Hole/Wormhole}

\noindent It is widely expected that the singularity problem in GR will ultimately be resolved by quantum gravity (QG) effects, although only a limited number of first-principles investigations currently substantiate this expectation~\cite{Dymnikova:1992ux,Dymnikova:2004qg,Ashtekar:2005cj,Bebronne:2009mz,Modesto:2010uh,Spallucci:2011rn,Perez:2014xca,Colleaux:2017ibe,Nicolini:2019irw,Bosma:2019aiu,Jusufi:2022cfw,Olmo:2022cui,Jusufi:2022rbt,Ashtekar:2023cod,Nicolini:2023hub}. Progress in this direction remains challenging, primarily because the fundamental QG framework is still unknown, thereby complicating the consistent incorporation of quantum effects. A widely adopted strategy consists in treating the unknown QG theory as an effective field theory~\cite{Donoghue:2012zc}. Within this paradigm, a particularly well-defined approach is provided by effective canonical QG, in which one attempts to quantize the canonical formulation of the effective theory, yielding a semiclassical description governed by an effective Hamiltonian constraint.

\noindent A central issue in this context concerns the conditions under which a given $(3+1)$ Hamiltonian model preserves full four-dimensional diffeomorphism covariance, i.e.\ corresponds to a generally covariant theory. While this requirement is straightforwardly enforced in the Lagrangian formulation through the construction of a generally covariant action, it becomes significantly more subtle in the Hamiltonian framework due to the explicit space-time decomposition. This so-called \textit{covariance issue} represents a fundamental challenge for effective canonical QG theories~\cite{Bojowald:2008gz,Tibrewala:2013kba,Bojowald:2015zha,Wu:2018mhg,Bojowald:2020xlw,Bojowald:2020unm,Gambini:2022dec,Bojowald:2022zog,Han:2022rsx,Giesel:2023hys}. A major advancement in addressing this problem was achieved in the work of Zhang, Lewandowski, Ma, and Yang~\cite{Zhang:2024khj}, where the analysis was carried out without imposing a specific gauge-fixing condition and focused on symmetry-reduced sectors relevant to loop quantum black hole models, i.e.\ within LQG.

\noindent In essence, Ref.~\cite{Zhang:2024khj} demonstrates that general covariance of an effective canonical QG theory imposes stringent constraints on the structure of the effective Hamiltonian. In particular, the Hamiltonian must be parametrized by an effective mass'' $M_{\text{eff}}$, which is required to satisfy a set of two covariance equations'' (see Eqs.~(7a,7b) of Ref.~\cite{Zhang:2024khj}). Distinct solutions to these equations correspond to different functional forms of $M_{\text{eff}}$, and therefore to inequivalent effective Hamiltonians $H_{\text{eff}}$, each defining a generally covariant effective QG theory by construction.

\noindent This formalism was subsequently applied to LQG-inspired models in Refs.~\cite{Zhang:2024khj,Zhang:2024ney}. Recall that LQG provides a background-independent, non-perturbative quantization of GR, in which the classical connection variables are replaced by holonomies through the polymerization procedure, leading to a fundamentally discrete space-time geometry (see e.g.\ Ref.~\cite{Ashtekar:2021kfp} for a review). Within the present framework, it is argued that holonomies effectively reduce to combinations of trigonometric functions of the extrinsic curvature. Consequently, polymerization can be understood as the replacement of the extrinsic curvature by suitable trigonometric functions, controlled by a new quantum (regularizing) parameter. This construction differs conceptually from the standard approach, in which polymerization is directly applied to the classical Hamiltonian constraint, typically yielding models whose general covariance is not guaranteed.

\noindent As a direct outcome of this methodology, Refs.~\cite{Zhang:2024khj,Zhang:2024ney} identified three independent solutions to the covariance equations, leading to three distinct effective Hamiltonian constraints. From these, three static, spherically symmetric (non-rotating) space-times were derived and shown to be closely connected to earlier loop quantum black hole models, thereby reinforcing the link between covariant effective canonical QG theories and LQG. These solutions have recently attracted considerable attention (see e.g.\ Refs.~\cite{Feng:2024sdo,Konoplya:2024lch,Liu:2024soc,Liu:2024wal,Malik:2024nhy,Heidari:2024bkm,Wang:2024iwt,Skvortsova:2024msa,Ban:2024qsa,Du:2024ujg,Lin:2024beb,Shu:2024tut,Liu:2024pui,Liu:2024iec,Paul:2025wen,Konoplya:2025hgp,Chen:2025ifv,Xamidov:2025oqx,Yang:2025ufs,Ai:2025myf,Wang:2025alf,Lutfuoglu:2025hwh,Chen:2025aqh,Sahlmann:2025fde,Zhang:2025ccx}). Of particular relevance for the present work is the third class of solutions derived in Ref.~\cite{Zhang:2024ney}, which constitutes a family parametrized by an integer $n$. The associated static, spherically symmetric line element, denoted by $ds^2_{(3)}$, is given by:
\begin{equation}
G(r) =\bar{f}_3^{(n)}\;\;,\;\;\; F(r)=\bar{\mu}_3^{-1} \left ( \bar{f}_3^{(n)} \right ) ^{-1}\;\;,\;\;\;H(r)=r^2,
\label{eq:ds23}
\end{equation}
where the function $\bar{f}_3^{(n)}(r)$ depends on the integer index $n \in \mathbb{Z}$ and is expressed as:
\begin{equation}
\bar{f}_3^{(n)}(r) = 1-(-1)^n\frac{r^2}{\ell^2}\arcsin \left ( \frac{2M\ell^2}{r^3} \right ) -\frac{n\pi r^2}{\ell^2},
\label{eq:f3}
\end{equation}
while the function $\bar{\mu}_3(r)$ is given by:
\begin{equation}
\bar{\mu}_3(r) = 1-\frac{4\ell^4M^2}{r^6}.
\label{eq:mu3}
\end{equation}
\noindent In Eqs.~(\ref{eq:f3},\ref{eq:mu3}), the parameter $\ell \propto \sqrt{\hbar}$ represents a quantum scale proportional to the Planck length, arising from the polymer quantization procedure and encoding the expected discrete structure of space-time at Planckian scales. The corresponding effective Hamiltonian constraint $H_{\text{eff}}^{(3)}$ is thus directly related to LQG.

\noindent The metric in Eq.~(\ref{eq:ds23}) asymptotically approaches the Schwarzschild solution only for $n=0$. In order to ensure phenomenological consistency, particularly with weak-field tests of gravity, we therefore restrict to the case $n=0$ in what follows. Under this assumption, the metric functions $f(r)$ and $g(r)$ appearing in Eq.~(\ref{eq:metric}) take the form:
\begin{align}
&G(r) = 1-\frac{r^2}{\ell^2}\arcsin \left ( \frac{2M\ell^2}{r^3} \right ) \label{eq:fr},
 \\
&F(r) = \left ( 1-\frac{4\ell^4M^2}{r^6} \right ) \left [ 1-\frac{r^2}{\ell^2}\arcsin \left ( \frac{2M\ell^2}{r^3} \right ) \right ] .
\label{eq:gr}
\end{align}
\noindent We refer to the resulting quantum-corrected geometry described by Eqs.~(\ref{eq:metric},\ref{falloffs},\ref{eq:gr}) as the “ZLMY black hole” (or more generally ZLMY space-time), following the initials of the authors of Ref.~\cite{Zhang:2024ney}. Notably, since $F(r) \neq G(r)$, the space-time is manifestly non-\textit{tr}-symmetric. Geometries of this type arise generically in quantum gravity approaches to singularity resolution~\cite{Modesto:2008im,Peltola:2008pa,Peltola:2009jm,Bianchi:2018mml,DAmbrosio:2018wgv}.

\noindent If the quantum parameter is in the range $0<\ell<\pi\sqrt{\pi}M^2/\sqrt{2} \approx 3.94 M^2$, the resulting geometry describes an asymptotically flat black hole with a horizon located at $r_H$, determined by the condition $F(r_H)=0$.
\noindent In this space-time, the classical singularity is replaced by a traversable wormhole structure with throat located at $r_{\min}=(2M\ell^2)^{1/3}$. Beyond the throat, the geometry extends into a Schwarzschild–de Sitter space-time with negative mass~\cite{Zhang:2024ney}. Importantly, no Cauchy horizon is present, rendering the solution free from mass inflation and, more generally, from perturbative instabilities. Regularity is explicitly confirmed by evaluating curvature invariants, including the Ricci scalar $R \equiv g^{\mu\nu}R_{\mu\nu}$, the Ricci tensor squared $R_{\mu\nu}R^{\mu\nu}$, and the Kretschmann scalar ${\cal K} \equiv R_{\mu\nu\rho\sigma}R^{\mu\nu\rho\sigma}$, explicitly reported in the Appendix of \cite{Calza:2025mwn}, remain finite for $r>r_{\min}$ and reduce to their Schwarzschild counterparts in the limit $\ell \to 0$. Accordingly, $\ell$ can be interpreted as a regularizing parameter. Overall, the ZLMY space-time describes a regular black hole geometry that resolves the classical singularity, avoids Cauchy horizons, and originates from a generally covariant effective canonical QG framework. This theoretical grounding provides a significantly more robust foundation compared to earlier phenomenological models, including our previous studies~\cite{Calza:2024fzo,Calza:2024xdh}.\\
\noindent Starting from a BH configuration, increasing the quantum parameter, the inner WH throat and the event horizon comes closer, coinciding for $\ell = \pi\sqrt{\pi}M^2/\sqrt{2}$. The solution is a wormhole for values of $\ell$ exceeding such limit.

\noindent From a phenomenological perspective, a key feature of the ZLMY black hole is that its temperature exceeds that of a Schwarzschild black hole of equal mass whenever $\ell \neq 0$. Although the temperature does not increase monotonically with $\ell$, it remains consistently higher than the Schwarzschild value $T_{\text{Sch}}=1/8\pi M$. In particular, the ratio $T/T_{\text{Sch}}$ reaches a maximum for $\ell \sim 3 M^2$, corresponding to an enhancement of approximately $25\%$ relative to the Schwarzschild case.

\noindent As for the CGVS metric, it is possible to compute its GBFs/transmission coefficients only through the numerically computed tortoise coordinates. Any attempt to obtain the tortoise coordinates using the analytic integral does not lead to results within {\tt GrayHawk} code.

\subsection{Morris-Thorne Wormholes }

The generic static, spherically symmetric traversable Morris--Thorne wormhole \cite{Morris:1988tu} is described by the line element
\begin{equation}
ds^2 = -e^{2\Phi(r)} dt^2 + \frac{dr^2}{1-\dfrac{b(r)}{r}} + r^2 \left(d\theta^2+\sin^2\theta\, d\phi^2\right).
\end{equation}

\noindent Here, $\Phi(r)$ is the redshift function, which must remain finite everywhere in order to avoid the presence of event horizons, thereby ensuring traversability. The function $b(r)$ is the shape function, which determines the spatial geometry of the wormhole and the location of the throat. The throat radius $r=\ell$ is defined by the condition
\begin{equation}
b(\ell)=\ell.
\end{equation}

\noindent For a traversable configuration, the flare-out condition must also hold at the throat,
\begin{equation}
b'(\ell)<1,
\end{equation}
while asymptotic flatness requires
\begin{equation}
\Phi(r)\to 0,
\qquad
\frac{b(r)}{r}\to 0,
\qquad
r\to\infty.
\end{equation}

\noindent This metric represents a two-way Lorentzian tunnel connecting two asymptotically flat regions of spacetime. In GR, sustaining such a geometry typically requires exotic matter violating the null energy condition.

\subsubsection{Ellis-Bronnikov Wormhole}

\noindent A particularly common and analytically convenient example of Morris-Thorne WH, is the zero-tidal-force Ellis-Bronnikov WH \cite{Ellis:1973yv,Bronnikov:1973fh} obtained by choosing
\begin{equation}\label{ZTFMT}
\Phi(r)=0,
\qquad
b(r)=\frac{\ell^2}{r},
\end{equation}
which yields the metric
\begin{equation}
ds^2 = -dt^2 + \frac{dr^2}{1-\dfrac{\ell^2}{r^2}} + r^2\left(d\theta^2+\sin^2\theta\,  \phi^2\right).
\end{equation}

\noindent This solution is horizonless, asymptotically flat, and possesses a throat at $r=r_0$. Since $\Phi(r)=0$, static observers experience no gravitational redshift, and tidal forces can be made arbitrarily small, making this geometry the standard prototype of a traversable wormhole.

\noindent The metric (\ref{ZTFMT}) may be rewritten adopting an intrinsic description as
\begin{equation}
    ds^2 = -dt^2 + d{\tilde r^2} + (\tilde r^2+\ell^2)\left(d\theta^2+\sin^2\theta\,  \phi^2\right)
\end{equation}

\subsection{Dadhich-Kar-Mukherji-Visser Wormhole}

We consider the static and spherically symmetric wormhole geometry introduced in Ref.~\cite{Dadhich:2001fu} and subsequently employed in several studies of wave propagation and compact-object phenomenology, including Ref.~\cite{Aneesh:2018hlp}. This lien element is part of the family of the self-dual Lorentzian WHs. The corresponding line element is given by
\begin{equation}
ds^{2}
=
-\left(\kappa + \lambda \sqrt{1-\frac{2M}{r}}\right)^{2} dt^{2}
+
\left(1-\frac{2M}{r}\right)^{-1} dr^{2}
+
r^{2}\left(d\theta^{2}+\sin^{2}\theta\, d\phi^{2}\right),
\label{SDLWmetric}
\end{equation}
where \(M\) is the mass parameter, while \(\kappa\) and \(\lambda\) are dimensionless constants characterizing deviations from the Schwarzschild solution.

\noindent This spacetime belongs to a family of solutions with vanishing Ricci scalar, \(R=0\), and may describe black holes, naked singularities, or traversable wormholes depending on the choice of parameters. In particular, for
\begin{equation}
\kappa = 0,
\qquad
\lambda = 1,
\end{equation}
Eq.~(\ref{SDLWmetric}) reduces to the Schwarzschild geometry. For nonvanishing \(\kappa\), however, the surface
\begin{equation}
r_{0}=2M
\end{equation}
is no longer an event horizon, but instead corresponds to the throat of a wormhole connecting two asymptotically flat regions.

\noindent Indeed, evaluating the temporal metric component at the throat yields
\begin{equation}
g_{tt}(r_{0})=-\kappa^{2},
\end{equation}
which remains finite and nonzero for \(\kappa \neq 0\). Consequently, the geometry is horizonless and, in principle, traversable.

\noindent The parameters \(\kappa\) and \(\lambda\) determine the redshift properties of the spacetime and control how closely the exterior geometry resembles that of a Schwarzschild black hole. For sufficiently small \(\kappa\), the object behaves as an ultra-compact black-hole mimicker, while preserving a wormhole throat in the interior region.

\noindent Owing to its analytical simplicity and its smooth interpolation between the Schwarzschild black hole and a traversable wormhole configuration, the metric (\ref{SDLWmetric}) provides a useful background for investigations of quasi-normal modes, gray-body factors, absorption spectra, and other scattering phenomena.

\subsection{Simpson-Visser and D'Ambrosio-Rovelli Wormhole}
Simpson-Visser and D'Ambrosio-Rovelli metrics were already present in the first version of $\tt GrayHawk$, and we refer the reader to their description in \cite{Calza:2025whq} for a brief introduction to such line elements.
Here we want to underline that $\tt GrayHawk$ $\tt v2$ is able to explore configurations of those metrics that were previously not accessible. Namely, for values of the regularizing parameter $\ell>2 M$ for which those metrics describe WHs. 

\subsection{Thin-Shell Wormholes}
The code permits the user to realize the thin-shell WH configuration of his preferences by allowing a cut and paste of metrics. Namely, it is possible to realize one of such a WH configuration by taking a WH/BH metric and choosing a cutting distance $\mathcal{T}>r_H$ or $\mathcal{T}>r_T$. The code will automatically copy and paste the solution to realize the corresponding thin-shell WH configuration.
As said, such a WH category is widely studied \cite{Poisson:1995sv,Visser:1989kg,Liempi:2024yjd,Sarkar:2026pjg,Eid:2024iza,Guo:2022iiy,Kokubu:2020lxs}. In fact, thin-shell provides one of the simplest and most tractable realizations of traversable wormholes, obtained by joining two spacetime manifolds across a hypersurface and concentrating the required exotic matter on an infinitesimally thin shell. This construction allows one to minimize and localize violations of the energy conditions, making it especially useful for investigating whether only small amounts of exotic matter are needed to sustain a wormhole. Their dynamics are also mathematically convenient, since stability under radial perturbations can often be reduced to the study of an effective potential for the throat radius. In addition, thin-shell wormholes can be constructed in a wide variety of gravitational theories beyond GR, which makes them valuable as comparative tools in modified gravity. Finally, because they may closely mimic black holes externally while lacking an event horizon, they are widely studied as possible black-hole alternatives with potentially observable signatures in lensing, quasi-normal ringing, scattering, and gravitational-wave echo phenomena.

\section{Numerical methodology and code description}

The second version of the code retains the previous version's organization, with two files. The primary file, $\tt GrayHawkv2.nb$, is designed to compute the gray-body factor or, more generally, transmission coefficient as a function of energy, expressed in units of the mass \( M \). The secondary file, $\tt CalibratorGHv2.nb$, serves as a supplementary tool to assist with calibration and parameter selection. The core structure of both codes is inherited from the previous version and has been enlarged to accommodate the possibility of obtaining transmission coefficients for WHs and to fully numerically obtain the tortoise coordinates for BHs.
To avoid repetition, we briefly outline the structure of the code and describe in detail only the newly introduced portion. We refer the reader to \cite{Calza:2025whq} for a full description of the unchanged portion of the code. We noted that the code oftentimes provides better results when the metric parameters are expressed as fractions rather than numbers with comma, e.g., prefer $\ell=3/2$ over $\ell = 1.5$.

\subsection{\tt GrayHawkv2.nb}
Here, we present the main tool employed to compute GBFs and the transmission coefficients of WHs. The program is implemented as a single, self-contained code module that can be executed through a single command. Once launched, the tool produces several outputs, including a plot of the functions $F(r)$ and $G(r)$ to verify the correct location of the horizon/throat, the plot of the geometric potential, a table listing the calculated GBF or transmission coefficient values, and the corresponding plot.

\noindent We maintained the code is structured into main sections, each divided into specific subsections as described below.
\begin{itemize}
    \item \textbf{Parameters} 
    \begin{itemize}
        \item \textit{\textbf{BH parameters}} We introduced the metric parameters relative to the metric described above and additionally pre-loaded in the code. It is also introduced here the radial distance of the cut $\mathcal{T}$ in the case of thin shell WH. 
        \item \textit{\textbf{Field parameters}}
        \item \textit{\textbf{Energy table}}
        \item \textit{\textbf{Smooth metric or cut\&paste/thin-shell WH}} The parameter named \newline $\tt 'SmoothOrNot'$ is added. It takes the values $1$ or $0$, selecting a smooth metric or thin-shell WH configuration, respectively. Those cases are implemented on dedicated $\tt If[]$ cycles. The thin-shell WH cycle is reported in red, and in this case, the WH throat value corresponds to the cut distance $\mathcal{T}$ by construction. 
        The smooth metric cycle is reported in blue, while black is left for the common parts. 
        \item \textit{\textbf{Tortoise coordinates analytic/numeric}}. The parameter named \newline $\tt 'TortoiseNumericDefinition'$ is added. It takes the values $0$ or $1$, selecting an analytic or numeric computation of the tortoise coordinates, respectively.
    \end{itemize}
    \item \textbf{The metric} According to the definition of the line element (\ref{eq:metric}), the metric functions $F(r)$, $G(r)$, and $H(r)$ are given in the specific subsections:
    \begin{itemize}
        \item \textit{\textbf{Definition of $F(r)$}}
        \item \textit{\textbf{Definition of $G(r)$}}
        \item \textit{\textbf{Definition of $H(r)$}}
    \end{itemize}
    In addition to the set of pre-compiled metrics previously available, we introduced BH and WH metrics, for some of which it is possible to obtain the tortoise coordinates solely in a numeric way. We also notice that there is the possibility to study the WH configuration of previously introduced metrics, such as the SV and the DR metrics. As for the previous code version, the user simply needs to uncomment the desired metric consistently across the three metric functions, and a custom line element can be introduced in this section. Altering the line element does not guarantee accurate results. Asymptotically flatness, providing a well-defined event horizon, conditions specified in (\ref{falloffs}) are necessary metric requirements preventing catastrophic code failure, but do not ensure accurate results.
    
    \begin{itemize}
        \item \textcolor{MidnightBlue}{\textit{\textbf{Horizon radius/throat definition and Assessment of Black Hole or Wormhole metric}} In this new subsection, the code assesses whether the metric and parameter choices describe a BH or a WH and identifies the horizon radius or radial throat value. Here, the user is asked to verify the correct location of the horizon/throat by checking the correct location of the largest root of $F(r)$ and $G(r)$ or checking that $\sqrt{H(0)}$ coincides with the throat value.}
        \item \textcolor{BrickRed}{\textit{\textbf{Warning to ensure that the metric is a thin shell WH}} The assessment is not possible for the choice $\tt 'SmoothOrNot=0'$. In such a case, given congenial choices of metric and cut distance, the line element should be a thin-shell WH, and the user is reminded of this with the phrase "Please make sure that the metric you consider together with the value of the cut $\mathcal{T}$ is a thin-shell wormhole metric".}
    \end{itemize}
    \item \textbf{Numerical inversion of tortoise coordinates}
    As in the previous version of the code, to have (\ref{SchEq}), it is necessary to determine \( V_s(r(r^*)) \), therefore, \( r(r^*) \). In this section, we implement a tripartite numerical procedure to obtain a continuous representation of \( r(r^*) \). This procedure is distinct depending on the nature of the considered object (BH or WH) and on the choice of $\tt 'TortoiseNumericDefinition'$.
    \textcolor{MidnightBlue}{\begin{itemize}
        \item In the case of smooth BH metric and $\tt 'TortoiseNumericDefinition'=0$, the code computes the inverted numerical tortoise coordinates starting with the analytic resolution of the integral defining the tortoise coordinates utilizing Mathematica's built-in $\tt{Integrate[] }$ function to compute the indefinite integral as described in \cite{Calza:2025whq}.
        \item In the case of smooth BH metric and $\tt 'TortoiseNumericDefinition'=1$, the code computes the inverted numerical tortoise coordinates starting with a numerical resolution of the integral defining the tortoise coordinates. Namely, the integral is performed by summing infinitesimal parallelograms and fixing the near-horizon value using an analytic approximation provided by a Taylor series truncated at first order near the horizon.
        \item In the case of smooth WH and $\tt 'TortoiseNumericDefinition'=0$, the code proceeds computing $r^*(r)$ as described in the first item of this list.
        \item In the case of smooth WH and $\tt 'TortoiseNumericDefinition'=1$, the code numerically computes the tortoise coordinates and proceeds to their inversion.
    \end{itemize}}
    \textcolor{BrickRed}{\begin{itemize}
        \item In the case of thin-shell WH and $\tt 'TortoiseNumericDefinition'=0$, the code proceeds computing $r^*(r)$ starting with the analytic resolution of the integral defining the tortoise coordinates utilizing Mathematica's built-in $\tt{Integrate[] }$ function to compute the indefinite integral as described in \cite{Calza:2025whq}.
        \item In the case of thin-shell WH and $\tt 'TortoiseNumericDefinition'=1$, the code numerically computes the tortoise coordinates and proceeds to their inversion.
    \end{itemize}}
    The code distinguishes those cases, using $\tt If[]$ cycles. The architecture within each cycle is similar and can be summarized in the subsections:
    \begin{itemize}
        \item \textbf{\textit{Analytic/Numerical  tortoise coordinate definition in case of BHs/WHs}} 
        \item \textbf{\textit{Numerical inversion}}
        \item \textbf{Inverted tortoise coordinates}
    \end{itemize}
    At the end of those cycles, the code defines the table of the inverted tortoise coordinates and its maximal and minimal values, and interpolates the table, obtaining a functional approximation. Since this last step, attention must be paid to preserve the characteristic of physical interest in the function of the inverted tortoise coordinates. Namely, the table of the inverted tortoise coordinates does not need to be overextended to the trivial regions at $r^* \rightarrow \pm \infty$. This way, the interpolation will provide an accurate result in the interesting region where the geometric potential differs from 0. The code is provided with a choice of compromise, aimed at maximizing the output accuracy for the different pre-loaded metrics. Choosing those values ad hoc for each different metric and field mode may improve the accuracy.
    
\item\textbf{The Geometrical potential}
    This section is identical to the previous version of the code, and it is structured as follows
    \begin{itemize}
        \item \textbf{\textit{Auxiliary functions}}
        \item \textbf{\textit{Geometric potential definition}}
        \begin{itemize}
            \item \textit{Geometric potential plot}
        \end{itemize}
    \end{itemize}
    \item \textbf{Solving for the GBF} Also, this section reads identically to the previous version of the code, but because it implements the core stage of the calculation, we discuss it here. The body of the code is the same, but since in the case of BHs and WHs the asymptotic values of the tortoise coordinates represent different regions, the code is computing the transmission coefficient of different problems. Namely, the scattering of waves propagating from a BH event horizon, scattering on the geometric potential and emerging in a distant region, or the scattering onto the geometric potential generated by a WH throat of a wave propagating between the two universes connected by the WH. A $\tt for[]$ loop is used to solve the scattering problem for each energy value in the table. The solution procedure follows the framework outlined previously: the radial equation~(\ref{SchEq}) is solved by imposing purely ingoing boundary conditions and normalizing the wave function; a region in the asymptotic far-field domain is then selected, where the numerical solution is sampled and fitted to the asymptotic expression (\ref{asympt+}); finally, the GBF is computed as $\Gamma=1/|b|^2$, where $b$ is determined from the far-field fit.
    
    \noindent At the conclusion of the $\tt for[]$ loop, the calculated values of \( \Gamma \) are compiled into a displayed table and subsequently plotted for visualization.
\end{itemize}

\noindent As default setting, the code parameters are all set to 0 with the exception of the BH mass, which must always be set to one, and the line element considered is the Schwarzschild one. The table of energies to be probed is made up of 100 equidistant values between 0 and 1 (in units such that $c=\hbar=G=M=1$).
The near and far regions are chosen according to the nature of the line element and the analytic/numerical choice of the user, depending on the minimum and maximum value of the sampler $r(r^*)$.
Finally the far portion of spacetime where to sample the function $Z_s$ is taken between the maximum value of the sampled inverted tortoise coordinate and $3/4$ of this distance.

\noindent Given such settings, the code provides all the output in roughly few seconds to more or less a minute on modern laptops.

\subsection{$\tt CalibratorGH.nb$}
\label{Calib}
In continuity with the previous version, a supplementary file has been released to assist users in accurately defining and sizing the near and far sampling regions to obtain a good reconstruction of the inverted tortoise coordinate, geometric potential (\ref{eq:potentials}), and consequently a reliable transmission coefficient.  

\noindent $\tt CalibratorGH$ $\tt v2$ focuses on the minimal and maximal energy values, \(\omega_{min}\) and \(\omega_{max}\), since once the scattering problem is well posed in such extrema, it is well posed for all intermediate steps. As for the main tool, this supplementary tool first informs the user on the nature of the chosen metric, providing the horizon/throat values, and asking the user to verify their correct location through plots. Therefore, the code provides the plot of the inverted tortoise coordinate together with the profile of \(V_s(r(r^*))\), allowing the user to assess whether the selected minimum and maximum values of \(r^*\) within the sampling interval are adequate for the potential to vanish in both asymptotic limits. The program then produces the profiles of \(Z_s(r^*)|_{\omega_{\min}}\) and \(Z_s(r^*)|_{\omega_{\max}}\), along with the corresponding values of \(\Gamma_s^{\,l}(\omega_{\min})\) and \(\Gamma_s^{\,l}(\omega_{\max})\). These diagnostics make it possible to examine two representative situations and verify that the numerical sampling parameters are suitable for the chosen energy interval.

\paragraph{Case 1: Sampling at \(\omega_{\min}\)}
If the sampling region is insufficient for \(\omega_{\min}\), the function \(Z_s(r^*)|_{\omega_{\min}}\) may fail to display the expected oscillatory behavior in the asymptotic far-field region (large \(r^*\)). This typically leads to an overestimation of \(\Gamma_s^{\,l}(\omega_{\min})\). In such cases, the far-away sampling domain should be enlarged so that the fit is performed farther from the horizon.

\paragraph{Case 2: Sampling at \(\omega_{\max}\)}
For \(\omega_{\max}\), inadequate sampling may instead appear as an oscillatory solution with increasing amplitude in the far-field region. This behavior generally signals that the near-horizon starting point has not been chosen sufficiently close to the horizon. As a consequence, the outward numerical integration begins from a suboptimal location, yielding an overestimate of \(\Gamma_s^{\,l}(\omega_{\max})\), which must remain smaller than \(1\). The issue can be mitigated by adopting a smaller near-horizon sampling distance.
\\
\\
\noindent In both cases, the parameter denoted by \(x_{\mathrm{far}}\) in the code must be selected with care so that the interval between \(x_{\max\infty}-x_{\mathrm{far}}\) and \(x_{\max\infty}\) lies entirely within the asymptotic far-away region. Moreover, the sampling frequency should satisfy the Nyquist--Shannon criterion to prevent aliasing effects.

\noindent When the energy range extends over several orders of magnitude, it may be advantageous to split the computation into multiple sub-ranges, assigning optimized sampling parameters to each sector. This strategy can substantially reduce computational cost while preserving numerical accuracy.

\section{Validation and Comparison}
In this section, we compare the results obtained with this code with some known results from the literature. Figs.~(1),~(2), and~(3-4) of this paper reproduce the transmission coefficients computed in the right panel of Fig.~(2) of \cite{Rosato:2025byu}, the lower panel of Fig.~(4) of \cite{Bao:2022iaz}, and the left panel of Figs.~(12),~(13) of \cite{Konoplya:2025hgp}, respectively. The first two cases account for thin-shell WHs constructed using the Schwarzschild metric. Ref.~\cite{Rosato:2025byu} locates the throat at $r_T=r_S(1+10^{-3})$ and $r_S(1+10^{-5})$, with $r_S=2M$ being the Schwarzschild radius, and takes into account the spin-2 field mode at $l=2$. While in Ref.~\cite{Bao:2022iaz}, the throat is set at $r_S(1+10^{-8})$, and it is considered a spin-2 field in the mode $l=5$. In both our computations for those two WHs, we chose to set $\tt TortoiseNumericDefinition=0$. Differently, Ref.~\cite{Konoplya:2025hgp} considers a ZLMY metric for different values of the parameter $\ell$ ($\ell=1, 2, 3, 3.91, 3.95, 4.1, 4.5, 12 $), and a spin-2 field of mode $l=3$. As previously noted, this metric does not allow analytical computation of the tortoise coordinates using our code. Therefore, we are forced to set $\tt TortoiseNumericDefinition=1$.

\noindent The code reproduces those transmission coefficients to such an accuracy that it is visually impossible to spot substantial discrepancies. We were able to compute the discrepancies between our computations and the original one since the authors of \cite{Bao:2022iaz,Rosato:2025byu, Konoplya:2025hgp} have been extremely available and helpful, providing the raw data of their computations. Specifically, we calculate the residual R as the absolute value of the difference, weighted over the average. Those R-factors are of the order of $10^{-4}$ in all the cases, except in the very narrow regions around extremely steep peaks, where even marginal shifts in the peak energy position enhanced by definition the R-factor. As an example, in Fig.~(5) we show the R-factor between our and \cite{Konoplya:2025hgp} computations in the case of ZLMY BH of $\ell=3.5$ and spin-2 wave mode $l=3$, blue dash-dotted line in Fig.~(4) and in Fig (13) of \cite{Konoplya:2025hgp}.

\begin{figure}[]
\centering
\includegraphics[width=0.8\linewidth]{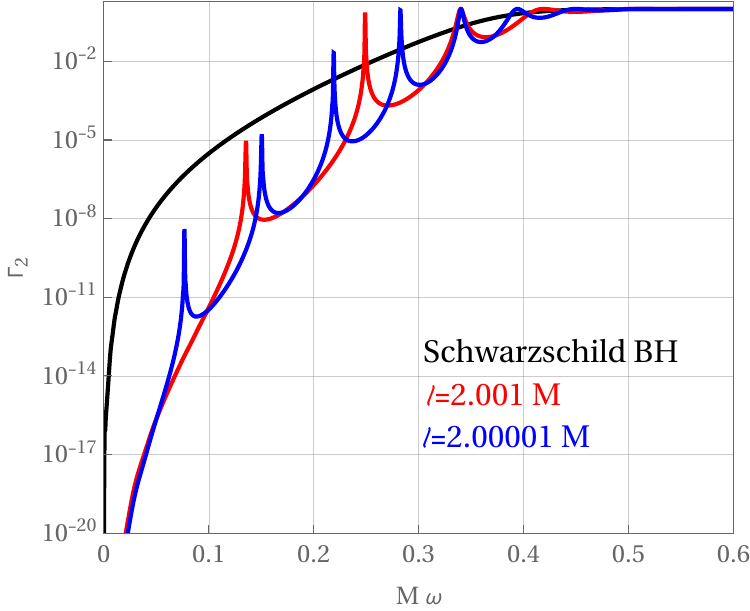}
\caption{Transmission coefficient of a field of spin-2 and mode $l=3$ scattering on a thin-shell WH constructed using the Schwarzschild metric and locating the throat at $r_T=r_S(1+10^{-3})$, red solid line, and $r_S(1+10^{-5})$, blue solid line, with $r_S=2M$ being the Schwarzschild radius. The black solid line represents the transmission coefficient of the same field and mode of a Schwarzschild BH.}
\end{figure}

\begin{figure}[]
\centering
\includegraphics[width=0.8\linewidth]{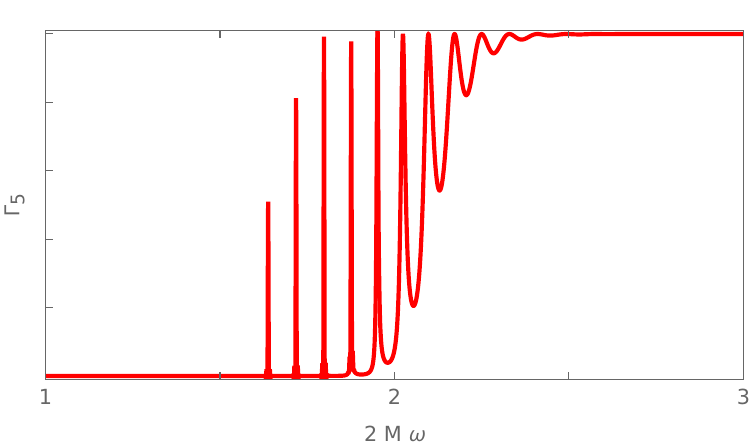}
\caption{Transmission coefficient of a field of spin-2 and mode $l=5$ scattering on a thin-shell WH constructed using the Schwarzschild metric and locating the throat at $r_T=r_S(1+10^{-8})$, red solid line, with $r_S=2M$ being the Schwarzschild radius.}
\end{figure}

\begin{figure}[]
\centering
\includegraphics[width=0.8\linewidth]{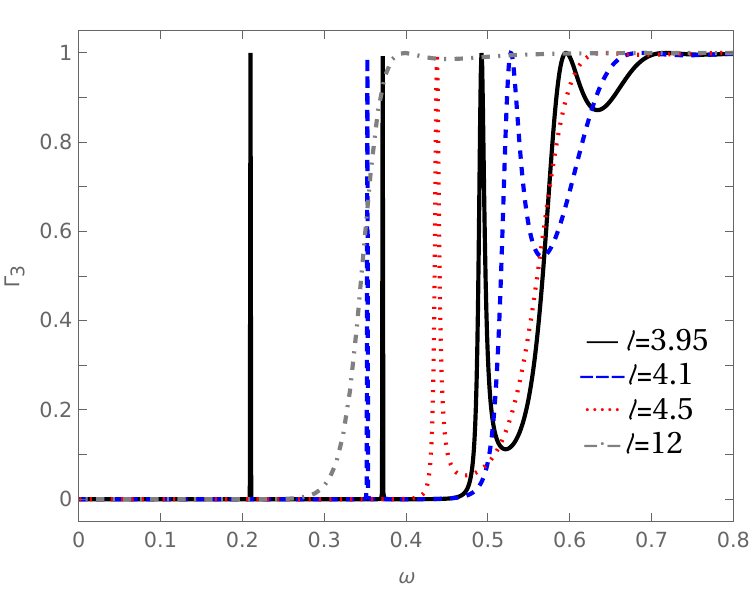}
\caption{Transmission coefficient of a field of spin-2 and mode $l=3$ scattering on ZLMY WH characterized by the regularizing parameter $\ell=3.95, 4.1, 4.5, 12$, in black solid line, blue dashed line, red dotted line, and gray dash-dotted line, respectively.}
\end{figure}

\begin{figure}[]
\centering
\includegraphics[width=0.8\linewidth]{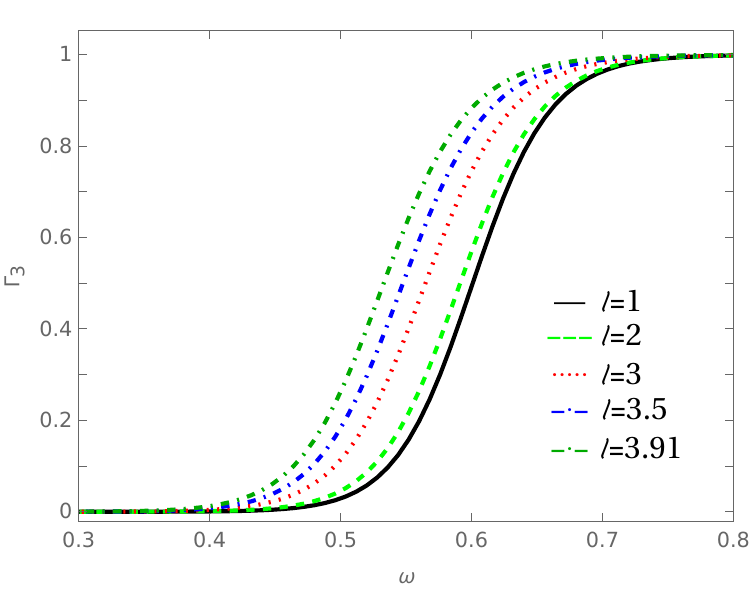}
\caption{Transmission coefficient of a field of spin-2 and mode $l=3$ scattering on ZLMY BH characterized by the regularizing parameter $\ell=1, 2, 3, 3.5, 3.91$, in black solid line, green dashed line, red dotted line, blue dash-dotted line, and dark green dash-sotted line, respectively.}
\end{figure}

\begin{figure}[]
\centering
\includegraphics[width=0.8\linewidth]{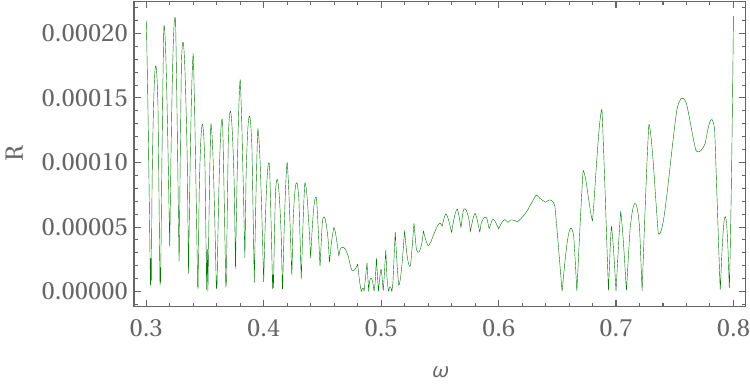}
\caption{Residuals between the gray-body factors of a wave of spin-2 and mode $l=3$ scattering on a ZLMY BH of $\ell=3.5$ computed with {\tt GrayHawk v2} and one showed in the right panel of Fig.~(13) of \cite{Konoplya:2025hgp}. The residuals are computed as the absolute value of the difference, weighted over the average.}
\end{figure}
\section{Conclusion}

\noindent In this work, we have presented \texttt{GrayHawk} \texttt{v2}, a substantial extension of the original Mathematica-based package introduced in Ref.~\cite{Calza:2025whq} for the computation of gray-body factors and related scattering observables in static, spherically symmetric compact-object spacetimes. The new release was motivated by two natural limitations of the first version: on the one hand, the restricted set of geometries for which the tortoise coordinate could be handled analytically within a practical workflow; on the other hand, the absence of horizonless compact objects among the supported backgrounds. Both issues become increasingly relevant in current gravitational research, where phenomenological studies often rely on non-standard metrics and on alternatives to classical black holes.

\noindent The first major improvement developed here is the implementation of a fully numerical treatment of the tortoise-coordinate integral. In many physically interesting metrics, the relation between the areal radius and the tortoise coordinate cannot be obtained in closed form, or its analytic expression is too cumbersome to be efficient. This represented an intrinsic bottleneck for the previous version of the code, since the tortoise mapping is a central ingredient in the Schr\"odinger-like formulation of the perturbation problem and therefore in the extraction of transmission and reflection coefficients. By introducing a robust numerical strategy for constructing, interpolating, and using the tortoise coordinate, \texttt{GrayHawk} \texttt{v2} considerably broadens the class of geometries that can be studied without requiring ad hoc analytic manipulations by the user. In this sense, the present release transforms a former technical limitation into a flexible computational asset.

\noindent The second major advance consists in extending the framework from black-hole spacetimes to traversable wormholes. This step is conceptually significant because wormholes, despite sharing several scattering features with black holes, possess no event horizon and therefore lead to qualitatively different conditions and wave dynamics. In particular, partial transmission across the throat and repeated internal reflections may generate resonant structures and late-time echo signals, making the associated transmission coefficients directly relevant for phenomenological investigations. By incorporating wormhole geometries into the package, \texttt{GrayHawk v2} now provides a unified environment in which black holes and horizonless compact objects can be analyzed within the same computational philosophy. This allows direct comparisons among models and facilitates systematic studies of how internal geometry affects observable scattering signatures.

\noindent The reliability of the new implementation has been tested against several benchmark results available in the recent literature. For thin-shell wormholes based on the Schwarzschild geometry, as well as for more involved metrics requiring a numerical tortoise construction, the transmission coefficients produced by the code agree with published results to a level such that discrepancies are visually negligible. These validations confirm both the correctness of the wormhole scattering module and the numerical stability of the new tortoise-coordinate procedure. They also show that the package can successfully handle cases in which analytical approaches are unavailable, precisely one of the main motivations for the present upgrade.

\noindent Throughout the development of \texttt{GrayHawk} \texttt{v2}, we have preserved the modular structure that characterized the original release. New metrics can be added with minimal intervention, numerical options can be adjusted according to the desired balance between speed and precision, and the separation between physical definitions and computational routines remains transparent to the user. We believe that this design philosophy is especially important for a community tool, since it allows researchers to adapt the package to new models, test conjectures rapidly, and integrate the code into broader phenomenological workflows.

\noindent It is worth emphasizing that, during the planning of this second release, one possible direction of development was the inclusion of higher-spin sectors, in particular the spin-$3/2$ field. Such an extension remains physically interesting, especially in contexts related to supergravity or beyond-standard-model scenarios. Nevertheless, after evaluating the range of potential applications, we considered the extension to wormhole spacetimes and the implementation of a general numerical tortoise treatment to be more timely and impactful improvements. These additions remove practical restrictions affecting a wide variety of metrics and simultaneously open the code to one of the most active areas in compact-object phenomenology. For this reason, they were prioritized in the present version. This choice should not be interpreted as excluding future developments: the inclusion of spin-$3/2$ fields remains a realistic and natural possibility for subsequent releases.

\noindent Looking ahead, several further upgrades appear promising. Among them are the extension to massive fields, rotating backgrounds, quasi-normal mode dedicated modules, improved resonance-search algorithms, automated parameter scans, and interfaces with waveform-analysis pipelines. In parallel, the addition of further exotic compact-object geometries could make the package increasingly useful in the interpretation of upcoming gravitational-wave and strong-field observational data. As numerical and observational precision continue to improve, tools capable of rapidly exploring broad families of models will become progressively more valuable.

\noindent In summary, \texttt{GrayHawk} \texttt{v2} significantly broadens the scientific scope of the original package. By overcoming the analytic dependence of the tortoise coordinate and by incorporating traversable wormholes into the scattering framework, it provides a more general, versatile, and future-oriented platform for the study of wave propagation on curved spacetimes. We hope that this public release will serve both as a practical instrument for current investigations and as a basis for further community-driven developments in black-hole, wormhole, and semiclassical gravity phenomenology.

\section*{Acknowledgments}
I am particularly grateful to Sunny Vagnozzi for the precious discussions, to all the friends and colleagues of the Theoretical Physics and Cosmology group of the University of Trento for their suggestions. I am thankful for the feedback provided by the \texttt{BlackHawk v3.0} developing team, Alexandre Arbay, Davide Pedrotti, Lea Malacher, and Yuber F. Perez-Gonzalez. A very warm thanks to Jo\~ao Rosa for his support over the years and for introducing me to the topics discussed in this paper.

\clearpage

\bibliography{GHII}

\end{document}